\documentclass[pmlr]{jmlr}%

\usepackage{booktabs, multirow, makecell, rotating}
\usepackage{float} %
\usepackage{xcolor,colortbl}
\usepackage{bbm}

\usepackage{todonotes}

\presetkeys{todonotes}{inline, size=\small, color=green!30}{}

\usepackage{longtable}%
 \usepackage{arydshln}

\usepackage[capitalise]{cleveref}

\usepackage[load-configurations=version-1]{siunitx} %

\makeatletter
\def\set@curr@file#1{\def\@curr@file{#1}} %
\makeatother

\theorembodyfont{\upshape}
\theoremheaderfont{\scshape}
\theorempostheader{:}
\theoremsep{\newline}

\jmlrvolume{252}
\jmlryear{2024}
\jmlrworkshop{Machine Learning for Healthcare}

\title[Needles in Needle Stacks]{Needles in Needle Stacks: Meaningful Clinical Information Buried in Noisy Waveform Data}

\makeatletter
\let\@fnsymbol\@arabic
\makeatother
\author{\Name{Sujay Nagaraj}\textsuperscript{*} \textnormal{\textsuperscript{1,2,3}}  \Email{s.nagaraj@mail.utoronto.ca} \\
\Name{Andrew J. Goodwin} \textnormal{\textsuperscript{4,5}}\Email{a.goodwin@sydney.edu.au} \\
\Name{Dmytro Lopushanskyy} \textnormal{\textsuperscript{4,6}} \Email{dmytro.lopushanskyy@cs.ox.ac.uk} \\
\Name{Danny Eytan} \textnormal{\textsuperscript{4,7}} \Email{danny.eytan@technion.ac.il} \\
\Name{Robert W. Greer} \textnormal{\textsuperscript{4}} \Email{robert.greer@sickkids.ca} \\
\Name{Sebastian D. Goodfellow} \textnormal{\textsuperscript{4,8}} \Email{sebastian.goodfellow@sickkids.ca} \\
\Name{Azadeh Assadi} \textnormal{\textsuperscript{4}} \Email{azadeh.assadi@sickkids.ca} \\
\Name{Anand Jayarajan} \textnormal{\textsuperscript{1,4}} \Email{anandj@cs.toronto.edu} \\
\Name{Anna Goldenberg} \textnormal{\textsuperscript{1,2,3,9,10}} \Email{anna.goldenberg@utoronto.ca} \\
\Name{Mjaye L. Mazwi} \textnormal{\textsuperscript{4,11}} \Email{mjaye.mazwi@seattlechildrens.org} \begin{flushleft}
\textnormal{\small{\noindent
\textsuperscript{1} Department of Computer Science, University of Toronto, Toronto, Canada\\
\textsuperscript{2} Department of Genetics \& Genome Biology, SickKids, Toronto, Canada\\
\textsuperscript{3} Vector Institute, Toronto, Canada\\
\textsuperscript{4} Department of Critical Care Medicine, SickKids, Toronto, Canada\\
\textsuperscript{5} School of Biomedical Engineering, University of Sydney, Sydney, Australia\\
\textsuperscript{6} Department of Computer Science, University of Oxford, Oxford, United Kingdom\\
\textsuperscript{7} Technion - Israel Institute of Technology, Haifa, Israel\\
\textsuperscript{8} Department of Civil \& Mineral Engineering, University of Toronto, Toronto, Canada\\
\textsuperscript{9} Canadian Institute for Advanced Research, Toronto, Canada\\
\textsuperscript{10} Department of Laboratory Medicine \& Pathobiology, University of Toronto, Toronto, Canada \\
\textsuperscript{11} Division of Critical Care Medicine, Seattle Children’s Hospital, Seattle, USA\\
\textsuperscript{*}\textnormal{Corresponding Author}}}
\end{flushleft}}
\begin{document}
\maketitle
\vspace*{-5mm}
\begin{abstract}
Central Venous Lines (C-Lines) and Arterial Lines (A-Lines) are routinely used in the Critical Care Unit (CCU) for blood sampling, medication administration, and high-frequency blood pressure measurement. Judiciously accessing these lines is important, as over-utilization is associated with significant in-hospital morbidity and mortality. Documenting the frequency of line-access is an important step in reducing these adverse outcomes. Unfortunately, the current gold-standard for documentation is manual and subject to error, omission, and bias. The high-frequency blood pressure waveform data from sensors in these lines are often noisy and full of artifacts. Standard approaches in signal processing remove noise artifacts before meaningful analysis. However, from bedside observations, we characterized a \emph{distinct} artifact that occurs during each instance of C-Line or A-Line use. These artifacts are buried amongst physiological waveform and extraneous noise. We focus on Machine Learning (ML) models that can detect these artifacts from waveform data in real-time - finding needles in needle stacks, in order to automate the documentation of line-access. We built and evaluated ML classifiers running in real-time at a major children's hospital to achieve this goal. We demonstrate the utility of these tools for reducing documentation burden, increasing available information for bedside clinicians, and informing unit-level initiatives to improve patient safety.
\end{abstract}

\section{Introduction}

In the Critical Care Unit (CCU), Central Venous Lines (C-Lines) and Arterial Lines (A-Lines) are used to sample blood, administer medications, and measure blood pressure in real-time. Over \emph{10 million} lines are placed each year in the USA alone \citep{scheer2002clinical, frasca2010prevention}. These lines are direct conduits from the bedside environment into a patient’s circulatory system. Despite their crucial role in the CCU, over-utilization of these lines is linked to a spectrum of adverse outcomes, most notably Central Line-Associated Bloodstream Infections (CLABSIs). In the USA alone, an estimated 48,600 - 80,000 CLABSI events occur in CCU patients every year \citep{o2002guidelines}. These are incredibly morbid events - patients who develop a CLABSI are more likely to have longer hospital stays, be readmitted, and die in hospital \citep{chovanec2021association, ziegler2015attributable}. Judicious use of vascular lines is essential in preventing \emph{unnecessary} morbidity and mortality for our critically ill patients. The classic Quality Improvement (QI) adage of \emph{``you cannot improve what you cannot measure’’} rings true, and monitoring of the amount, duration, and frequency of line access is a key factor in preventing adverse outcomes \citep{frasca2010prevention, misset2004continuous}. Unfortunately, the current gold standard for monitoring involves manual entry by the bedside clinician accessing the line. This process is known to introduce errors, omissions, and inherent biases \citep{bahl2024reliability}.

\begin{figure}[t]
  \centering 
  \includegraphics[width=\textwidth]{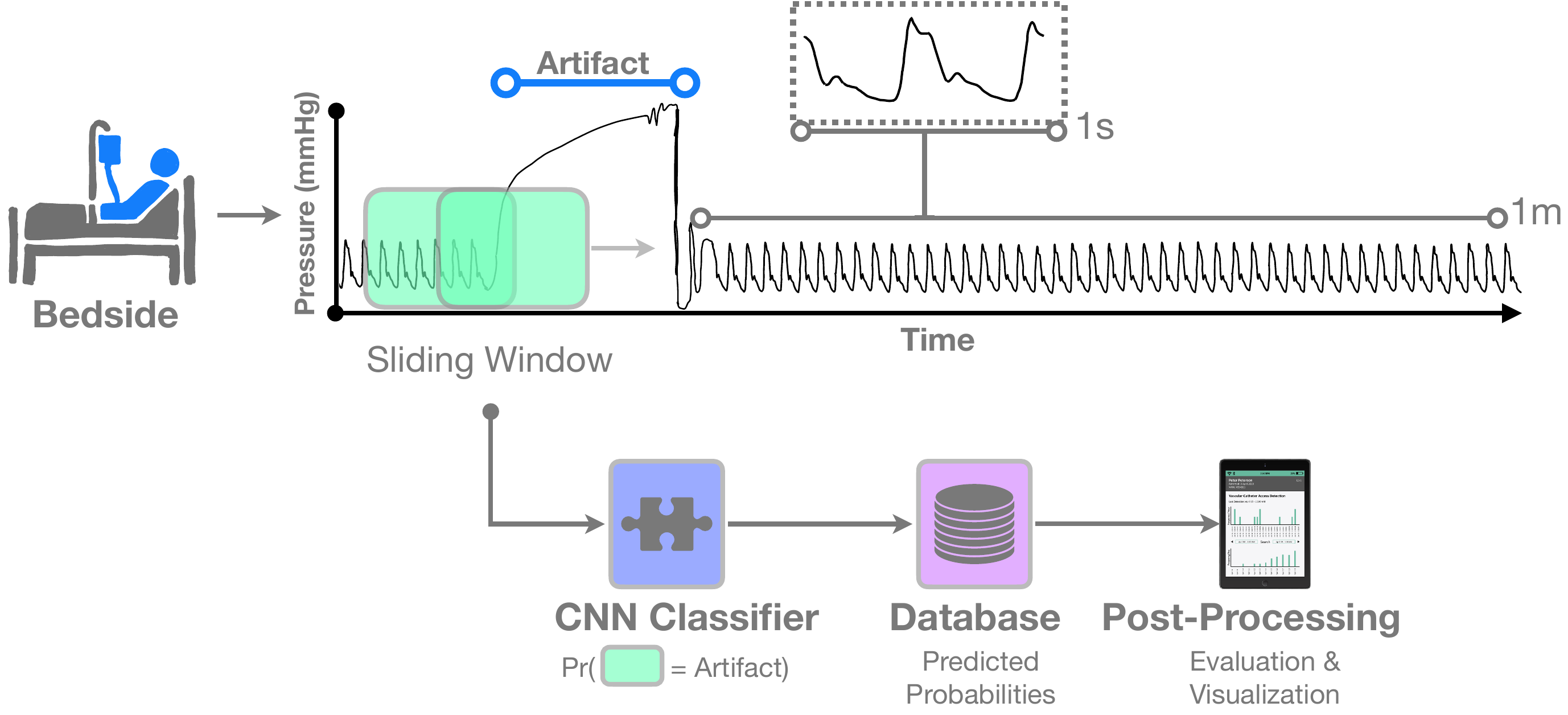} 
  \caption{High-frequency blood pressure data from vascular lines in patients provide important information about patient state. Vascular line-access generates artifacts (as shown) - note the size and scale compared to physiological waveform data. We employ a sliding window approach for training a CNN classifier and running real-time inference. }
  \label{fig:figure1} 
\end{figure}

We propose that automating the detection of line-access events in real-time with highly accurate models can serve as an important clinical tool to improve patient care and safety. In order to achieve this, we leveraged a distinct artifact in blood pressure waveform data, occurring during each line-access event (\cref{fig:figure1}). Detection of these events in real-time proves a technical challenge as it requires precise identification of relatively infrequent events in vast quantities of high-frequency data - identifying needles in needle stacks. We developed, rigorously evaluated, and deployed Machine Learning (ML) classifiers designed to accomplish this task, thereby automating the monitoring of line access.

In this paper, we document the path of these algorithms from the clinic to GPU and back again. We first demonstrate that the noise in high-frequency waveforms actually carries meaningful clinical information. We then build simple models with exceptional performance to detect these needles in needle stacks. Though contemporary approaches which \emph{remove} artifacts \citep{cao2006simple, islam2021signal, son2018automated} are still useful, we propose that making sense of noise artifacts can yield additional dividends. We identify a specific clinical use-case where artifact detection can inform initiatives to prevent line over-utilization. We take this a step further and showcase other downstream clinical benefits of using our tool from the perspective of various clinician end-users. The insights we describe in this paper reframe the importance of noise in waveform data and how simple models can be deployed to net outsized returns from these artifacts.

\newpage

\subsection*{Generalizable Insights about Machine Learning in the Context of Healthcare}

\begin{enumerate}
\item \textbf{Noise itself represents meaningful clinical information}. The noise in high-frequency waveform data is often \emph{overlooked} or \emph{removed} in traditional analyses. Though this is important for certain analyses, we demonstrate a specific case where noise artifacts actually capture information about bedside interventions. Documenting these interventions in real-time by detecting these artifacts can help improve patient safety.
\item \textbf{Simple models yield outsized returns with good design}. Well-performing, \emph{simple} models that exploit the noise in waveform data may help prevent the adverse impacts of vascular over-utilization. Additionally, we show how these models can benefit multiple clinician end-users to impact care in a multitude of unexpected ways. We demonstrate this with a real-world deployment at a major children's hospital.
\end{enumerate}

\section{Related Work}

\paragraph{Clinical Context}
Vascular lines can be placed in arteries or veins: arterial lines (A-Lines) and central venous lines (C-Lines) both have different purposes as well as frequencies of use. Notably, A-Lines are more likely to be present but are less frequently used \citep{gershengorn2014variation}. They are predominantly used for sampling arterial blood (oxygenated blood from the heart) as well as monitoring Arterial Blood Pressure (ABP). C-Lines are inserted less frequently, but when they are, they are associated with increased acuity. Once in place, C-Line access occurs much more frequently than A-Lines. These lines are routinely used for sampling (deoxygenated blood to the heart), monitoring Central Venous Pressure (CVP), and they can also be used for medication administration (i.e., infusions).

Over-utilization of both A-Lines and C-Lines is correlated with increased morbidity and mortality for hospitalized patients \citep{o2002guidelines, pronovost2006intervention}. In order to minimize adverse events at the individual patient- and unit-level, awareness of utilization patterns is essential \citep{frasca2010prevention, misset2004continuous}. Current means to document utilization are restricted to manual annotation by bedside nurses - e.g., a nurse would document the event after taking a blood draw for laboratory tests. However, manual documentation is subject to error, omission, and bias \citep{bahl2024reliability}. For example, instances where the line was accessed for troubleshooting are not documented and therefore missed in overall reporting. At the unit-level, a clinician audits the number of times documentation occurred and is responsible for reporting overuse. However, audits do not happen in real-time, so patients are continually at risk until a clinician deliberately looks at the data.

\paragraph{Noise-generating Process}

During each line-access event, a characteristic artifact is generated in the waveform data (\cref{fig:figure1}). This artifact is easily recognized by any clinician who has worked with these lines because the process that generates it is very well characterized \citep{li2009artificial}. This artifact forms due to the physical process of accessing a vascular line. Initially, a stopcock is closed to allow access between the vascular circulation and the syringe used by the clinician. Blood pressure builds up and is read by the transducer as blood continues to flow against this closed port. Pressure is then released once the stopcock is reopened following the clinical intervention (e.g., blood draw). In the pressure waveform, this shows up as a rise and subsequent drop in blood pressure, which is what generates the characteristic shape of the artifact in question. Though there are many other types of noise found in waveform data, the specific noise-generating process of the line-access artifact and its distinct shape allow us to consider how to detect it automatically.

\paragraph{Signal Processing \& Anomaly Detection}
Various noise and artifacts are scattered throughout high-frequency waveform data. Artifacts arise from many sources including clotting, transducer flushing, patient movement, non-invasive cuff inflation, and line access \citep{mcghee2002monitoring}. Existing paradigms focus on identifying and \emph{removing} artifacts before analysis \citep{khan2022optimized, mannan2018identification, nizami2013implementation, son2018automated}. Both traditional signal processing techniques and more modern ML-based anomaly detection approaches are used to this end \citep{edinburgh2021deepclean, son2018automated, sun2006signal}. For example, \citet{edinburgh2021deepclean} use generative approaches for self-supervised artifact rejection in CCU blood pressure waveform data.

Other ML algorithms for high-frequency waveform data in the CCU focus on decision-support using predictive models \citep{hatib2018machine, johnson2016machine, mollura2021novel}. These models aim to perform important tasks such as providing early warnings for cardiac arrest or sepsis. However, widespread use remains bottlenecked by low interpretability, lack of infrastructure outside academic settings, and lofty prediction tasks with high variability in performance \citep{d2022identifying, norrie2018challenge, tonekaboni2019clinicians}. In contrast, our approach has inherent interpretability as we are detecting specific waveform artifacts that clinicians know arise from a well-established noise-generating mechanism. The mechanism is independent of patient physiology, thus not impacted by inter-patient variability that may plague other predictive approaches (e.g., sepsis prediction). Lastly, because of the characteristic shape of these artifacts in contrast to normal physiologic waveforms, this represents a `low-hanging fruit' classification task. Going against the grain, by detecting artifacts instead of removing them, can provide valuable clinical insights and may also be more amenable to clinical adoption.

\section{Methods}

\subsection{Dataset}
\label{methods:dataset}
We used blood pressure waveform data sampled at 125Hz (samples/second) collected from patients in the CCU at The Hospital for Sick Children (SickKids) in Toronto, Canada. Our institution has been collecting such waveform data continuously from all bedspaces in the CCU since 2016. We began with a bedside clinical observer manually labeling line-access events and their corresponding waveform artifacts. The observer documented each line-access event, and we linked these 1:1 to the artifact in question - of particular note is the distinct, \emph{``sharkfin''}-like appearance of the artifact in contrast with the normal blood pressure waveform (\cref{fig:figure1}). Labeling yielded a training dataset comprising approximately $1800$ unique A-Line access events recorded between 2016 and 2018. We performed a similar procedure to label C-Line access events between January and June 2022. The labeling was done across complete waveform data from each of these periods, which ensured that whatever was not labeled is implicitly non-artifact (i.e., negative class). Formally, we constructed a training dataset of size $n$ with fixed-length window of size $W \in \mathbb{N}$, $\mathcal{D} = \{(x_{1:W},y)_i\}_{i=1}^n$. Where $x_{1:W} \in \mathbb{R}^W$, and $y \in \{0,1\}$ with $1$ representing a positive label for artifact. Specific dataset sizes for A-Line and C-Line data across all window sizes can be found in \cref{Appendix::Datasets}.

\subsection{Modeling}
Envisioning a simple binary classification task using 1-dimensional waveform data, we opted for a Convolutional Neural Network (CNN) binary classifier, using different fixed-length window sizes $W$ as a key hyper-parameter. The CNN classifier, $f: \mathbb{R}^W \rightarrow [0, 1]$, maps the input window of the waveform to a sigmoid score, which approximates the posterior probability of a window containing an artifact:  $f(x_{1:W}) \approx \textnormal{Pr}(y=1 \mid x_{1:W})$. Models were initially trained on non-overlapping windows from both classes: labeled artifact intervals ($y=1$) and non-artifact \emph{or} unrelated artifact intervals ($y=0$). The deliberate addition of artifacts \emph{not} associated with line-access events allowed us to penalize the model for detecting artifacts we are not interested in (e.g., motion artifacts).

We trained a separate model for A-Lines and C-Lines across multiple sliding-window sizes $W \in \{ 1250, 2500, 3750, 7500, 12500, 15000 \}$. Because the waveform is sampled at 125Hz, this set corresponds to window sizes of 10s, 20s, 30s, 60s, 100s, 120s, respectively. Instances were normalized to zero-mean and unit-variance based on the preceding ten minutes of waveform data. This allowed us to accommodate for patient-specific variability in mean blood pressure and to preserve the size difference of the artifacts compared to normal physiological waveform (\cref{fig:figure1}). A CNN for each $W$ was fit using mini-batch SGD with Cross-Entropy loss and early-stopping monitoring for validation loss. Hyper-parameter selection for different CNN parameters (e.g., kernel size, dropout, etc.) optimized for validation AUC using the KerasTuner \citep{omalley2019kerastuner}.

\paragraph{Streaming Inference} During streaming inference (both retrospective and prospective), models were deployed using a sliding window approach (\cref{fig:figure1}). Here, a fixed length window is moved along the time-series at fixed increments to generate predictions across the entire waveform. The CNN classifier outputs the probability of artifact presence for each overlapping window. Given a $T$ length waveform, $x_{1:T}$, and a window of width $W$, where $W < T$, the process of sliding the window along the series with a step-size of $S$ results in splitting the waveform into overlapping segments: these segments are defined as $x_{1+kS:(1+kS)+W-1}$ for $k = 0, 1, 2, \dots, \left\lfloor\frac{T-W}{S}\right\rfloor$, where $T$ is the length of the time series, and $k$ represents the step index. The floor function $\left\lfloor \cdot \right\rfloor$ ensures that $k$ does not extend the window beyond the end of the series. These windows were then passed into the model for inference. Predicted probabilities and the associated time-stamps for each window were stored in a SQL database along with metadata about the bedspace they originated from before post-processing and analysis. Additional details on model architecture, training, tuning, as well as code to reproduce experiments can be found in \cref{Appendix::Implementation}.

\subsection{Post-processing Techniques}
\label{postprocess}
\begin{figure}[t]
  \centering 
  \includegraphics[width=\textwidth]{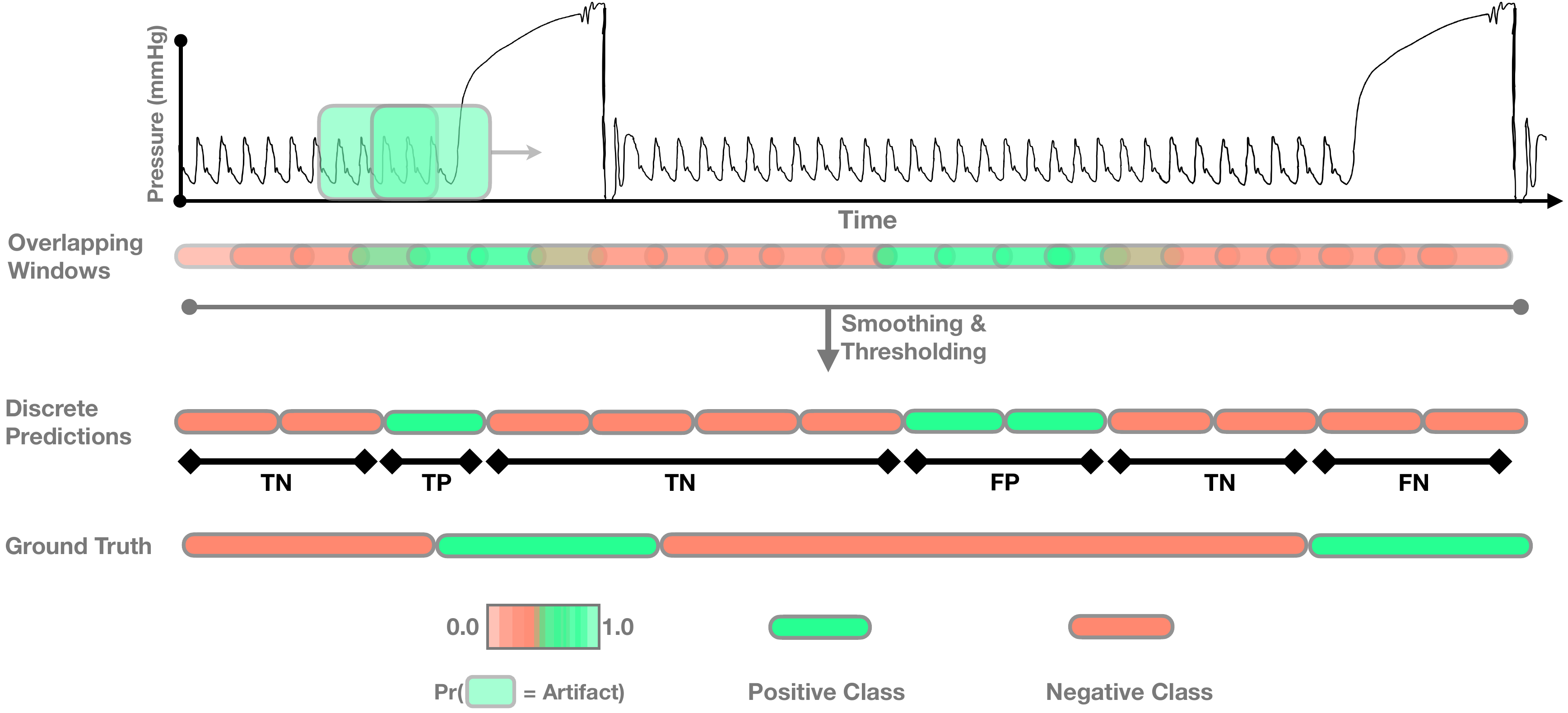} 
  \caption{Post-processing of predicted probabilities from overlapping sliding-windows are smoothed using a Gaussian convolution and thresholded to generate a distinct prediction for each non-overlapping window. Examples of how individual predictions were compared to ground-truth intervals.}
\label{fig:post_process} 
\end{figure}

The sliding-window technique generated predicted probabilities for overlapping regions of waveform - e.g., a 60s interval of time may have multiple distinct predictions associated with it depending on the step-size. In order to generate a single classification for each interval of time, we applied a Gaussian convolution to average over the predicted probabilities from overlapping sliding-windows (\cref{fig:post_process}). Here, we center a Gaussian kernel over a window of interest, and compute the mean of the predicted probabilities of all windows that overlap that region, weighted by the degree of overlap and distance from the center of the current window of interest: Formally, the Gaussian kernel centered at window $t_c$ is defined as:
\begin{align*}
G(t; t_c, \sigma) = \frac{1}{\sigma \sqrt{2\pi}} e^{-\frac{(t-t_c)^2}{2\sigma^2}}
\end{align*}
where $t_c$ is the center of the current window, and $\sigma$ is the standard deviation of the Gaussian distribution.

For a given window centered at $t_c$, we compute a weighted average of the predicted probabilities over all overlapping windows. The predicted probability for the window centered at $t_c$, after applying the Gaussian weighting, is given by:
\begin{align*}
A(t_c) = \frac{\sum_{t=t_c-W/2}^{t_c+W/2} G(t; t_c, \sigma) \cdot f(x_{t:t+W})}{\sum_{t=t_c-W/2}^{t_c+W/2} G(t; t_c, \sigma)}
\end{align*}
These post-processing techniques allowed us to efficiently condense overlapping predictions to discrete ground-truth intervals \cref{fig:post_process}.

\subsection{Evaluation}
\label{sec::evaluation}
\subsubsection{Static Evaluation}
Static evaluation was done on the held-out non-overlapping windows from our original training dataset (20\% held-out) for each model. We denote this as static evaluation, as it was not done in a sequential fashion. We report test-set performance for a variety of metrics relevant for such an anomaly detection task: Accuracy, Precision, Recall / True Positive Rate (TPR), and False Positive Rate (FPR). Static evaluation allowed us to evaluate for overfitting during model training and identify how well our model can detect the distinct characteristics of the artifact in question. However, this evaluation technique is ill-equipped to generalize to continuous, streaming inference.

\subsubsection{Retrospective Evaluation}
\label{subsec::retro_eval}

In order to evaluate the model in a manner consistent with a real-world streaming deployment, we held-out continuous intervals of labeled data from one bedspace (1 year for A-Line and 6 months for C-Line data). Each dataset contains continuous waveform data across all patients occupying that bed during this time interval. Because our training datasets were curated to achieve favorable class-balance, it is (intentionally) out-of-distribution (\cref{methods:dataset}). Retrospective evaluation ensured that the models would be evaluated in as close to a real-world setting as possible to see if static performance metrics translate to a streaming context - subject to the same class-balance and frequency of real-world line-access artifacts, as well as being tested on unrelated artifacts that arise in the waveform naturally. This technique allowed us to properly validate model performance prior to prospective deployment in a manner consistent with the real-world. Evaluation in the static setting where discrete labels are associated with each example and prediction is straightforward - it is easy to generate a confusion matrix. However, translating this to a continuous, streaming setting required some thought, as it is not entirely clear how continuous \emph{intervals} of predictions should be compared to (possibly) imprecisely-labeled ground-truth intervals. We adopted the following classification schema to evaluate streaming predictions:

\begin{enumerate}
\item \textbf{True Positive} - positive \emph{predicted} window overlapping with positive \emph{ground-truth} interval.
\item \textbf{False Positive} - positive \emph{predicted} window  sufficiently far (30s) from the nearest positive \emph{ground-truth} interval.
\item \textbf{True Negative} - negative \emph{predicted} window overlapping with negative \emph{ground-truth} interval OR partial overlap with a \emph{ground-truth} positive interval, provided a different predicted window correctly classified that event.
\item \textbf{False Negative}: negative \emph{predicted} window  overlapping with negative \emph{ground-truth} interval, with no other predicted windows correctly classifying that event.
\end{enumerate}
    
\cref{fig:post_process} provides a visual representation on how predictions were classified with examples of each type of classification. We allowed for some flexibility in False Positive evaluation because the labels themselves are not precise, with the labeled start and stop time of artifacts potentially deviating from the true times. As such, a positive predicted window is flagged as a False Positive only if it is \emph{sufficiently far} (i.e., 30s) from a True Positive interval. For evaluating True and False Negatives, we also do not penalize if single windows err on a true artifact, provided the artifact was picked up by different windows. This design choice was due to the understanding that accurately detecting the presence of artifacts should be prioritized over accurately detecting the exact start and stop of these events. Additionally, the labelled start and stop times for each event is not precise and subject to error. Evaluating against such a noisy label may not yield useful results.

\begin{figure}[t]
  \centering 
  \includegraphics[width=\textwidth]{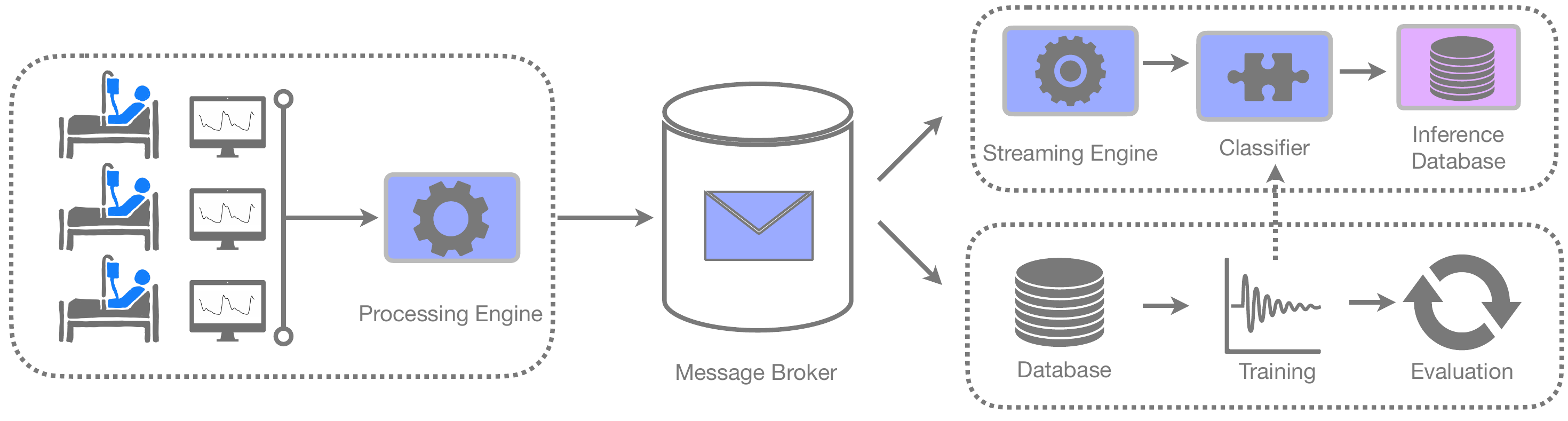} 
  \caption{Physiological waveform streaming architecture. Real-time data is processed and sent to a message broker. Data for training and model evaluation is stored in a proprietary database. For inference, streaming waveform data is pre-processed and fed to a trained model in a sequential, batched fashion. Model predictions are stored in a SQL database back-end before post-processing and analysis.}
  \label{fig:stream} 
\end{figure}

\subsection{Threshold Calibration}
\label{calibration}

A unique binary prediction at each interval is typically generated by thresholding the predicted probability of each window, $\hat{y}_{t_c} = \mathbbm{1}[ A(t_c) > 0.5 ]$. However, binary classifiers trained on static datasets may not necessarily be calibrated for a streaming setting. This discrepancy arises because crucial parameters, such as class balance, differ between the static training set and the dynamic streaming environment. 

The threshold calibration process in our study is conducted as follows: we first split the Retrospective Evaluation dataset (\cref{subsec::retro_eval}) into $n=5$ evaluation folds of equal duration (i.e., each equal duration fold is a continuous period of waveform intervals). For each evaluation fold, we use the remaining $n-1$ folds as the calibration set. For each calibration set, we perform threshold calibration which involves finding the optimal classification threshold that maximizes a specific evaluation metric - we optimize for the $\textnormal{F}_1$-score on the calibration set. This process helps to ensure that the binary classification decisions are optimized for metrics of clinical interest. $\textnormal{F}_1$-score was selected to find an optimal trade-off between Precision and Recall in downstream predictions using a grid-search of thresholds from $0.5$ to $1$. This optimal threshold is then used to classify the intervals in the evaluation set and calculate the evaluation metrics outlined in \cref{sec::evaluation}. Given the natural trade-off between Precision and Recall, future practitioners may prioritize different metrics based on the clinical application. %

\subsection{Prospective Deployment}
After retrospective evaluation, we engineered a streaming pipeline capable of ingesting data from all bedspaces within the CCU for simultaneous real-time inference across the unit (\cref{fig:stream}). This infrastructure adeptly manages the dynamic nature of the clinical environment - ranging from patient admissions and discharges to disconnected lines. Our deployment infrastructure is general to other models in the CCU at our institution, with the use-case described in this paper being an important real-world test of the infrastructure.

After receiving the necessary ethical clearance, we embarked on a non-interventional, prospective trial aimed at validating the model’s efficacy in real-time \citep{kwong2022silent, tonekaboni2022validate}. This initiative is supported by a collaborative effort with clinician end-users, ensuring the models' integration into clinical practice is both seamless and impactful. The models are currently deployed in the CCU, and we specifically analyzed predictions from November 2023 to February 2024 after applying the aforementioned post-processing techniques in \cref{postprocess}. In the following sections, we present prospective results of A-Line and C-Line models trained on 60s window-size.

\section{Results}

\definecolor{fgood}{HTML}{BAFFCD}
\definecolor{fbad}{HTML}{FFC8BA}
\definecolor{funk}{HTML}{FFFFE0}

\newcommand{\goodvalue}[1]{\cellcolor{fgood}\textnormal{#1}}
\newcommand{\badvalue}[1]{\cellcolor{fbad}\textnormal{#1}}

\begin{table*}[h]
\centering
\resizebox{\textwidth}{!}{
\begin{tabular}{clccccc}
& Window Size (s) & Accuracy (\%) $\uparrow$ & Precision (\%) $\uparrow$ & TPR/Recall (\%) $\uparrow$  & FPR (\%) $\downarrow$ \\
\cmidrule(lr){2-6}

\multirow{6}{*}{\rotatebox[origin=c]{90}{A-Line}}
& 10 & \goodvalue{99.4$\pm$0.2} & \goodvalue{97.8$\pm$0.9} & \goodvalue{97.9$\pm$1.1} & \goodvalue{0.4$\pm$0.2}  \\

& 20 & \goodvalue{99.4$\pm$0.1} & \goodvalue{96.8$\pm$0.7} & \goodvalue{99.1$\pm$0.2} & \goodvalue{0.5$\pm$0.1}  \\

& 30 & \goodvalue{99.4$\pm$0.4} & \goodvalue{96.2$\pm$1.7} & \goodvalue{99.2$\pm$0.7} & \goodvalue{0.6$\pm$0.4}  \\

& 60 & \goodvalue{99.4$\pm$0.4} & \goodvalue{95.7$\pm$3.5} & \goodvalue{99.0$\pm$1.2} & \goodvalue{0.6$\pm$0.4} \\

& 100 & \goodvalue{99.1$\pm$0.6} & \goodvalue{99.0$\pm$2.0} & \goodvalue{92.4$\pm$4.4} & \goodvalue{0.1$\pm$0.3} \\

& 120 & \goodvalue{98.8$\pm$0.6} & \badvalue{88.4$\pm$7.1} & \goodvalue{94.8$\pm$4.7}  & \goodvalue{0.9$\pm$0.5} \\
\cdashline{2-6}

\multirow{6}{*}{\rotatebox[origin=c]{90}{C-Line}}
& 10 & \goodvalue{93.9$\pm$0.4} & \goodvalue{95.7$\pm$0.5} & \goodvalue{96.8$\pm$0.5}  & \badvalue{19.7$\pm$1.5}  \\

& 20 & \badvalue{80.6$\pm$1.0} & \goodvalue{98.6$\pm$0.8} & \badvalue{74.1$\pm$1.7} & \goodvalue{2.6$\pm$1.6}  \\

& 30 & \goodvalue{91.8$\pm$1.0} & \goodvalue{94.5$\pm$1.0} & \goodvalue{90.2$\pm$1.8}  & \goodvalue{6.2$\pm$1.3}  \\

& 60 & \goodvalue{96.5$\pm$0.9} & \goodvalue{92.6$\pm$1.8} & \goodvalue{97.4$\pm$1.2}  & \goodvalue{3.9$\pm$0.9}  \\

& 100 & \goodvalue{97.1$\pm$0.7} & \goodvalue{94.2$\pm$2.8} & \goodvalue{97.3$\pm$2.1} & \goodvalue{2.9$\pm$1.2}  \\

& 120 & \goodvalue{97.2$\pm$1.1} & \goodvalue{93.8$\pm$2.5} & \goodvalue{95.2$\pm$2.6} & \goodvalue{2.1$\pm$0.9}  \\
\bottomrule
\end{tabular}
}
\caption{Comparison of Static Evaluation metrics for A-Line and C-Line models across various window sizes. Each metric is reported as mean value $\pm$ standard deviation across 10 folds. Results in \colorbox{fgood}{\textbf{green}} exceed our target-performance ($>90\%$ for Acc, Prec, TPR and $<10\%$ for FPR), all other results are in  \colorbox{fbad}{\textbf{red}}.}
\label{tab:static_eval}
\end{table*}

\begin{table*}[h]
    \centering
    \resizebox{\textwidth}{!}{
    \begin{tabular}{clccccc}
    & Window Size (s) & Accuracy (\%) $\uparrow$ & Precision (\%) $\uparrow$ & TPR/Recall (\%) $\uparrow$ & FPR (\%) $\downarrow$ \\
    \cmidrule(lr){2-6}
    
    \multirow{6}{*}{\rotatebox[origin=c]{90}{A-Line}}
    & 10 & \goodvalue{99.9$\pm$0.1} & \goodvalue{98.1$\pm$0.9} & \goodvalue{91.7$\pm$9.9}  & \goodvalue{0.0$\pm$0.0} \\
    
    & 20 & \goodvalue{99.9$\pm$0.1} & \goodvalue{98.1$\pm$0.9} & \goodvalue{91.7$\pm$9.9} & \goodvalue{0.0$\pm$0.0}  \\
    
    & 30 & \goodvalue{99.9$\pm$0.1} & \goodvalue{96.6$\pm$2.0} & \goodvalue{95.6$\pm$6.9}  & \goodvalue{0.0$\pm$0.0}  \\
    
    & 60 & \goodvalue{99.8$\pm$0.1} & \goodvalue{98.2$\pm$1.9} & \badvalue{81.9$\pm$13.1}  & \goodvalue{0.0$\pm$0.0} \\
    
    & 100 & \goodvalue{99.0$\pm$0.3} & \goodvalue{100.0$\pm$0.0} & \badvalue{17.4$\pm$13.2}  & \goodvalue{0.0$\pm$0.0}  \\
    
    & 120 & \goodvalue{98.9$\pm$0.5} & \badvalue{89.4$\pm$13.7} & \badvalue{28.6$\pm$15.7} & \goodvalue{0.0$\pm$0.0}  \\
    \cdashline{2-6}
    
    \multirow{6}{*}{\rotatebox[origin=c]{90}{C-Line}}
    & 10 & \goodvalue{99.7$\pm$0.1} & \goodvalue{92.3$\pm$5.2} & \badvalue{88.6$\pm$5.0}  & \goodvalue{0.1$\pm$0.0} \\
    
    & 20 & \goodvalue{99.7$\pm$0.1} & \goodvalue{93.7$\pm$4.5} & \goodvalue{91.1$\pm$2.8}  & \goodvalue{0.1$\pm$0.0} \\
    
    & 30 & \goodvalue{99.6$\pm$0.5} & \goodvalue{92.1$\pm$5.7} & \badvalue{86.1$\pm$3.2}  & \goodvalue{0.0$\pm$0.0}  \\
    
    & 60 & \goodvalue{99.2$\pm$0.6} & \goodvalue{92.7$\pm$5.2} & \goodvalue{97.8$\pm$2.5}  & \goodvalue{0.0$\pm$0.0}  \\
    
    & 100 & \goodvalue{98.4$\pm$0.9} & \badvalue{88.9$\pm$8.2} & \badvalue{78.1$\pm$6.0}  & \goodvalue{0.6$\pm$0.2}  \\
    
    & 120 & \goodvalue{98.5$\pm$1.1} & \badvalue{88.9$\pm$7.8} & \badvalue{80.3$\pm$4.7} & \goodvalue{0.6$\pm$0.2} \\
    \bottomrule
    \end{tabular}
    }
    
    \caption{Comparison of Retrospective Evaluation metrics for both A-Line and C-Line models across various window sizes. Each metric is reported as mean value $\pm$ standard deviation across 5 folds. Results in \colorbox{fgood}{\textbf{green}} exceed our target-performance ($>90\%$ for Acc, Prec, TPR and $<10\%$ for FPR), all other results are  \colorbox{fbad}{\textbf{red}}.}
    \label{tab:retro_eval}
\end{table*}

\begin{figure}[t]
  \centering 
  \includegraphics[width=\textwidth]{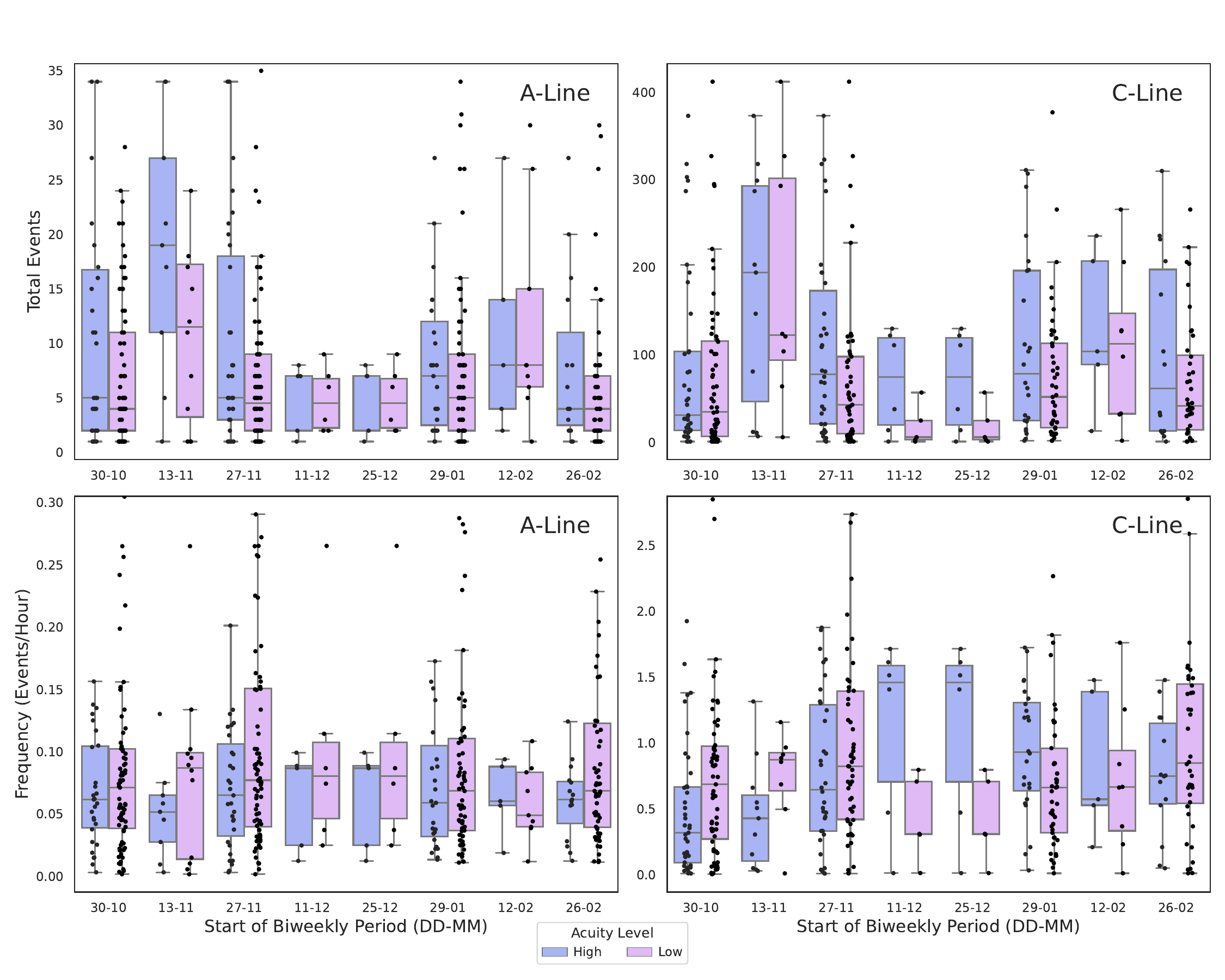} 
  \caption{Total line-access events and Frequency of line-access events (events/hour) from model predictions running real-time between November 2023 and February 2024. We compute each metric at the patient-level and plot them for each bi-weekly period (i.e., each point represents the metric for a single, unique patient). Results are stratified by the inferred acuity-level of each patient, as certain bedspaces are reserved for higher-acuity patients. We observe seasonal trends in line-access that can help inform QI initiatives targeting unit-level line-access rates. We also note a higher-burden of line-access frequency among those with higher-acuity.}
 \label{fig:over_time} 
\end{figure} 

\begin{figure}[t]
  \centering 
  \includegraphics[width=\textwidth]{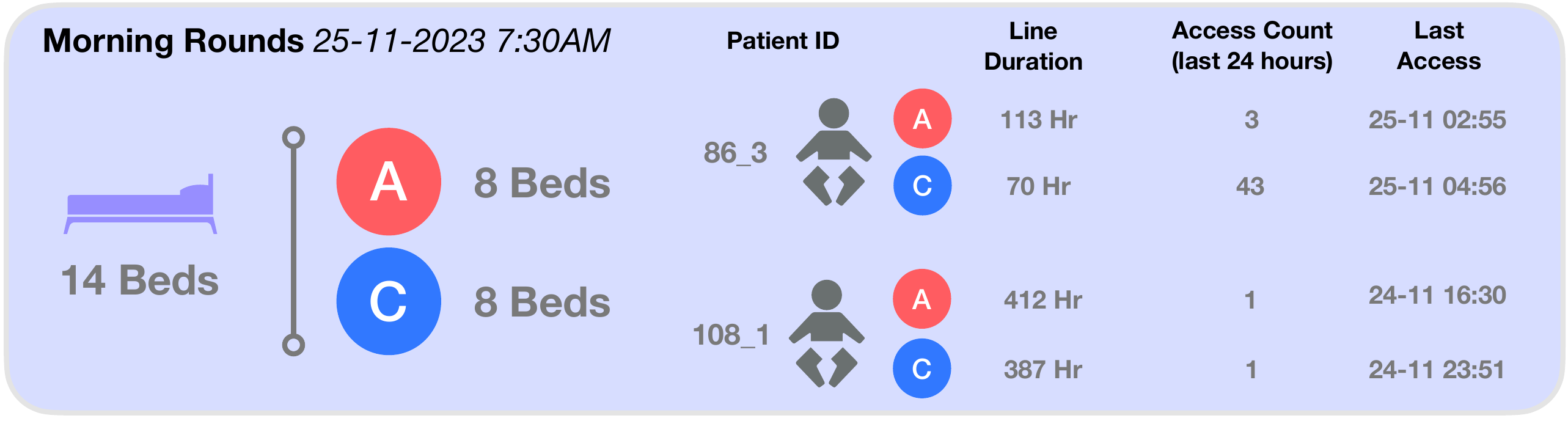} 
  \caption{Mock-up report for CCU Morning Rounds on status of A-Lines and C-Lines at the unit- and patient-level. Patient-level summary statistics reflect which lines each patient has, total duration of each line, access count in the last 24 hours, and time of last access. Values displayed come from real model inferences on real patients from a de-identified date.}
  \label{fig:unit_stats}
\end{figure} 

\subsection{On the Utility of Noise in High-Frequency Sensor Data}

A core insight from this work is that noise in high-frequency physiological waveform data can be leveraged to glean clinically-relevant insights. This stemmed from a bedside observation that specific artifacts are associated with each line-access event. As discussed earlier, this goes against conventional wisdom in signal processing and analysis of high-frequency data. We demonstrate a specific and realizable clinical use-case leveraging data that are typically discarded via traditional signal processing paradigms (e.g., high-pass filtering, outlier removal, etc.).

High-frequency waveform data explicitly encode information about bedside interventions. Because each event represents some sort of clinical intervention (e.g., medication administration, or blood sampling, etc.), waveform data contains embedded information about clinical judgement for that patient - with higher utilization frequency potentially reflecting higher acuity patients. Stepping back, each line-access artifact represents a clinical decision taken about that patient at a given time. Changes in the frequency of these events not only reflects unit-level utilization patterns, but changes in the clinical management of individual patients.

Lastly, we believe the output of our models can provide yet another temporal measure to track in the CCU. In an observational study, \cite{agniel2018biases} found that the presence and timing of blood tests, which are typically drawn from vascular lines, are much more predictive of mortality than the results of the tests themselves. Alongside information about patient vitals (e.g., heart rate, respiration, etc.), real-time line-access frequency can serve as a valuable input for training more complicated models that predict patient mortality \citep{muralitharan2021machine, tonekaboni2018prediction}. 

\subsection{On Simple Models for High-Impact Tasks}
A key part of our approach is the deliberate choice of using simple model architectures to perform high-impact tasks - this approach allowed us to train models with exceptional performance with little time spent in the model development and optimization stages. In \cref{tab:static_eval}, we achieve near perfect performance with respect to Accuracy, Precision, TPR, and FPR. The simplicity of the task allowed us to maintain high standards for model performance. For example, in \cref{tab:static_eval} and \cref{tab:retro_eval}, we highlight models that achieved our target-performance of  $>90$\% (or $<10$\% for FPR). 

Our choice of a CNN-based classifier required few design choices, can be trained efficiently without the need for GPUs, using out-of-the-box tools, and also demonstrates excellent performance on both Static and Retrospective Evaluation (\cref{tab:static_eval} and \cref{tab:retro_eval}, respectively).

\subsection{On the Importance of Rigorous Retrospective Evaluation}
Many ML models deployed in healthcare settings experience sharp drop-offs in performance compared to the metrics evaluated during early evaluation phases \citep{nestor2019feature, zhang2022shifting}. The early phase tends to rely on \emph{Static Evaluation} - often involving curated datasets, which may not reflect the reality of distribution shifts during streaming-inference. Because of the relatively low real-world frequency of line-access events, it makes sense to try to utilize balanced training datasets so that our models can learn a good representation of the artifact of interest. However, during inference time, the models are searching for needles within needle stacks of noisy waveform data. In order to estimate what real-world performance would look like, we deployed our A-Line and C-Line models retrospectively on continuous regions of held out training data in a streaming fashion. This allowed us to expose the models to the (lack of) class balance found in real-world data, as well as other non-relevant artifacts that would be encountered. For both A-Line and C-Line models, we see performance drop-offs when evaluating these metrics in a retrospective fashion, however performance generally passes our standards (\cref{tab:static_eval} vs. \cref{tab:retro_eval}). Of note, when it comes to deciding an optimal window size, what works well in the Static setting does not necessarily carry over to the Retrospective setting. As such, we believe the latter evaluation setting is the ideal way to tune such hyper-parameters before a prospective deployment. A reason why window size is important may be due to the variable duration of the line-access artifacts - artifacts can be 10s or last several minutes. We can see that with larger window sizes performance tends to drop-off, likely because we are missing the smaller artifacts that take up a relatively small proportion of a window (\cref{tab:retro_eval}).

\subsection{On Designing for Clinician End-Users}
In our prospective deployment of this model, we performed real-time inference on A-Line and C-Line data from all beds in the CCU between November 2023 to February 2024. This amounted to over 150,000,000 individual model inferences. After post-processing, we identified over 25,000 individual line-access events across over 400 unique patients admitted in the CCU. Here we provide specific clinical use-cases for our models that are designed for our end-users, highlight how we can improve patient care in a multitude of ways. 

\paragraph{Documentation Burden}
Few ML models deployed are built and explicitly designed to reduce documentation workload, especially for bedside nursing staff \citep{o2023artificial}. In contrast, our models are designed to reduce the (error-prone) documentation of line-access events. This can yield dividends on clinician stakeholder buy-in, particularly from nurses. In consultation with our nursing stakeholders, we found that each line-access event adds roughly $1$ minute of extra documentation time, our models detected over $1900$ line-access events in the month of January 2024, indicating that automating line-access detection can save around $30$ hours of  documentation time on a monthly basis for our nursing staff. This adds up over the course of a year, and is time that can be reallocated towards patient care. This is particularly meaningful during a period where nurses are overworked in under-staffed and under-resourced healthcare settings \citep{ford2023overworked, suran2023overworked}. Additionally, the simplicity of our model, and the known cause-and-effect relationship between line-access and artifact-generation among clinicians can help build trust with these models \citep{tonekaboni2019clinicians}. 

\paragraph{Clinical Care}
Detecting line-access events in real-time can also have direct impacts on clinical care. As we discussed, insights about line-access frequency are essential in mitigating adverse events associated with their use. Each morning, a clinical team \emph{rounds} on all the CCU patients to get a birds-eye assessment on each patient's state. In \cref{fig:unit_stats}, we provide a mock-up of what information our models' predictions can allow the clinical team to have on-hand during these `Morning Rounds'. This mock-up showcases real-time model predictions on real patients from a specific (de-identified) date, at the time that rounds typically take place (7:30AM). Clinicians can monitor the total number of beds with A-Lines or C-Lines, and also see how these are distributed amongst the patients. Lastly, summary statistics representing the total time the line has been in, number of accesses in the last 24 hours, and the time of last access can also be valuable information. We can see from the figure that one of the patients shown (86\_3) has had substantially more line-access events in the past 24 hours than the other. This communicates previous and current clinical interventions as well as clinician judgement about a patient's state - this patient has required \emph{more} interventions in the past 24 hours. In fact, this patient is also in a bedspace typically associated with higher acuity in our CCU. Additionally, patient 108\_1 has had both an A-Line and C-Line for 16 days with only 1 access in the past 24 hours - bedside providers can use this information to evaluate the risks and benefits of vascular lines in real-time.

\paragraph{Quality Improvement (QI)}
Our last use-case reflects how aggregate statistics of unit-level access frequency can help guide QI initiatives to mitigate over-utilization of vascular lines. In \cref{fig:over_time}, we showcase A-Line and C-Line access events identified over each biweekly period from November 2023 to February 2024. We observe seasonal fluctuations in both total number of events as well as frequency - with notable drops in utilization as patient volume goes down during the winter holidays. In our CCU, there are particular bedpsaces that are associated with higher-acuity patients due to their need for increased monitoring. When we further stratify these results based on bedspace acuity, we see how the burden of line-access events and frequency  differ at the patient-level and over time.

\section{Discussion}
\label{Discussion}
In this work we challenge the idea that noise in high-frequency data is devoid of meaningful information. Starting with a bedside observation, we were able to identify a characteristic artifact buried in noisy waveform data that signalled line-access utilization. The identification of line-access artifacts within blood pressure waveform data has unveiled a rich source of information pertinent to clinical interventions, offering novel insights into patient-state and risk characterization. Through the development and deployment of models capable of identifying these artifacts in real-time, this study explores the paradigm of leveraging the \emph{noise} in physiological waveform data for tangible clinical benefit. We also demonstrate that the direct integration of our model into the clinical workflows of end-users has the potential to enhance patient safety which can generalize to any clinical setting using vascular lines.

\paragraph{Limitations}
Our work addresses how line-access artifacts encode valuable clinical information. However, it is important to characterize what information is buried in \emph{other} noise artifacts. For example, artifacts caused by patient movement obscure physiological waveform signals but may also encode information about patient discomfort or distress. Though we perform a rigorous retrospective evaluation, it is important to verify model performance prospectively as well. We are currently undertaking a convenience-sample based approach to compare model predictions to actual line-access events documented by bedside observers. Our approaches heavily rely on the post-processing techniques we outline in \cref{fig:post_process}, however, we still require rigorous evaluation on how well these techniques fair compared to other approaches - though this is out of the scope of this paper. Another limitation is scalability, though we demonstrate that noise encodes meaningful clinical information, our framework is limited to CCUs with the ability to store and analyze waveform data. These models may only be feasible in a small subset of academic hospitals. Though as technological advancements permeate the field of Critical Care Medicine, these disparities may be reduced. Lastly, ongoing monitoring and evaluation will be required to ensure models continue to operate as intended in the face of unavoidable distribution shifts.

\paragraph{Acknowledgements}
S.N.was financially supported by: Ontario Graduate Scholarships, SickKids ResTraComp PhD Award, CIHR Vanier CGS Award, Mr. Robert and Francine Ruggles MD/PhD Innovation Award. This work was also supported by the William G. Williams Directorship for Analytics at The Hospital for Sick Children (M.L.M.), a Varma Family Chair (A.G.), and a CIFAR AI Chair (A.G.).

\bibliography{references}

\newpage
\appendix

\section{Datasets}
\label{Appendix::Datasets}
\subsection{Static Evaluation}

\begin{table}[h]
\scalebox{0.5}
\centering
\begin{tabular}{lcccccc}
\toprule
\multirow{2}{*}{\textbf{$W$ (s)}} & \multicolumn{2}{c}{\textbf{Train}} & \multicolumn{2}{c}{\textbf{Val}} & \multicolumn{2}{c}{\textbf{Test}} \\
                                      & \textbf{A-Line} & \textbf{C-Line} & \textbf{A-Line} & \textbf{C-Line} & \textbf{A-Line} & \textbf{C-Line} \\
\midrule
10 & \begin{tabular}[c]{@{}c@{}}$n^{+} = 6601$\\ $n^{-} = 39511$\end{tabular} & \begin{tabular}[c]{@{}c@{}}$n^{+} = 10653$\\ $n^{-} = 3365$\end{tabular} & \begin{tabular}[c]{@{}c@{}}$n^{+} = 2250$\\ $n^{-} = 13094$\end{tabular} & \begin{tabular}[c]{@{}c@{}}$n^{+} = 3585$\\ $n^{-} = 1075$\end{tabular} & \begin{tabular}[c]{@{}c@{}}$n^{+} = 2202$\\ $n^{-} = 13101$\end{tabular} & \begin{tabular}[c]{@{}c@{}}$n^{+} = 5112$\\ $n^{-} = 1116$\end{tabular} \\
20 & \begin{tabular}[c]{@{}c@{}}$n^{+} = 3125$\\ $n^{-} = 19284$\end{tabular} & \begin{tabular}[c]{@{}c@{}}$n^{+} = 5196$\\ $n^{-} = 3293$\end{tabular} & \begin{tabular}[c]{@{}c@{}}$n^{+} = 975$\\ $n^{-} = 6500$\end{tabular} & \begin{tabular}[c]{@{}c@{}}$n^{+} = 1485$\\ $n^{-} = 1038$\end{tabular} & \begin{tabular}[c]{@{}c@{}}$n^{+} = 1081$\\ $n^{-} = 6446$\end{tabular} & \begin{tabular}[c]{@{}c@{}}$n^{+} = 2771$\\ $n^{-} = 1089$\end{tabular} \\
30 & \begin{tabular}[c]{@{}c@{}}$n^{+} = 1948$\\ $n^{-} = 12638$\end{tabular} & \begin{tabular}[c]{@{}c@{}}$n^{+} = 3752$\\ $n^{-} = 3235$\end{tabular} & \begin{tabular}[c]{@{}c@{}}$n^{+} = 656$\\ $n^{-} = 4235$\end{tabular} & \begin{tabular}[c]{@{}c@{}}$n^{+} = 1158$\\ $n^{-} = 1030$\end{tabular} & \begin{tabular}[c]{@{}c@{}}$n^{+} = 635$\\ $n^{-} = 4198$\end{tabular} & \begin{tabular}[c]{@{}c@{}}$n^{+} = 1236$\\ $n^{-} = 1043$\end{tabular} \\
60 & \begin{tabular}[c]{@{}c@{}}$n^{+} = 730$\\ $n^{-} = 5948$\end{tabular} & \begin{tabular}[c]{@{}c@{}}$n^{+} = 1778$\\ $n^{-} = 2995$\end{tabular} & \begin{tabular}[c]{@{}c@{}}$n^{+} = 271$\\ $n^{-} = 1955$\end{tabular} & \begin{tabular}[c]{@{}c@{}}$n^{+} = 584$\\ $n^{-} = 1043$\end{tabular} & \begin{tabular}[c]{@{}c@{}}$n^{+} = 256$\\ $n^{-} = 1993$\end{tabular} & \begin{tabular}[c]{@{}c@{}}$n^{+} = 490$\\ $n^{-} = 968$\end{tabular} \\
100 & \begin{tabular}[c]{@{}c@{}}$n^{+} = 304$\\ $n^{-} = 2244$\end{tabular} & \begin{tabular}[c]{@{}c@{}}$n^{+} = 937$\\ $n^{-} = 2857$\end{tabular} & \begin{tabular}[c]{@{}c@{}}$n^{+} = 86$\\ $n^{-} = 749$\end{tabular} & \begin{tabular}[c]{@{}c@{}}$n^{+} = 140$\\ $n^{-} = 912$\end{tabular} & \begin{tabular}[c]{@{}c@{}}$n^{+} = 99$\\ $n^{-} = 741$\end{tabular} & \begin{tabular}[c]{@{}c@{}}$n^{+} = 456$\\ $n^{-} = 928$\end{tabular} \\
120 & \begin{tabular}[c]{@{}c@{}}$n^{+} = 188$\\ $n^{-} = 2223$\end{tabular} & \begin{tabular}[c]{@{}c@{}}$n^{+} = 619$\\ $n^{-} = 2748$\end{tabular} & \begin{tabular}[c]{@{}c@{}}$n^{+} = 69$\\ $n^{-} = 743$\end{tabular} & \begin{tabular}[c]{@{}c@{}}$n^{+} = 271$\\ $n^{-} = 914$\end{tabular} & \begin{tabular}[c]{@{}c@{}}$n^{+} = 61$\\ $n^{-} = 742$\end{tabular} & \begin{tabular}[c]{@{}c@{}}$n^{+} = 302$\\ $n^{-} = 911$\end{tabular} \\
\bottomrule
\end{tabular}
\caption{Training / Static Evaluation dataset sizes for window sizes ($W$(s)) across A-Line and C-Line, indicating positive class ($n^{+}$) and negative class ($n^{-}$) counts.}
\label{table:window_sizes_subcategories}
\end{table}

\subsection{Retrospective Evaluation}
\paragraph{A-Line} Data for A-Line retrospective evaluation consisted of 548 manually-labelled line-access artifacts from a single bedspace spanning July 2016 - March 2018.

\paragraph{C-Line} Data for C-Line retrospective evaluation consisted of 407 manually-labelled line-access artifacts from two bedspaces spanning January 2022 - June 2022.

\section{Implementation Details}
\label{Appendix::Implementation}
\paragraph{Models} CNN models were built in Python using the Keras Library \citep{chollet2015keras}. Both A-Line and C-Line models shared the same architecture: two convolution layers, batch normalization, dropout, average pooling, linear layer, dropout, and a final linear layer. Models were trained using the Adam optimizer \citep{kingma2014adam} and a binary cross-entropy loss. Hyperparameters (number of convolution filters, learning rate, and dropout rates) were tuned using the KerasTuner library \citep{omalley2019kerastuner} optimizing for Validation AUC.

\paragraph{Retrospective and Prospective Deployment} All models were deployed using a proprietary streaming engine for waveform data pre-processing and inference ~\citep[see e.g.,][for more details]{Goodwin2020, Jayarajan2021}. The engine collects data from bedside devices, applies filters, and sends them to a message broker. The message broker can then be consumed in real-time by an ML algorithm or sent to an archival time-series database \cref{fig:stream}. 

Code to train models and recreate experiments can be found in \href{https://github.com/sujaynagaraj/line_labelling_public.git}{a public repository}. Unfortunately, we are unable to release data due to privacy considerations at our institution and cannot release prospective deployment code at this time as it would identify the institution and authors.
\end{document}